# Evaluating AI-Enabled deception vulnerability amongst Sub-Saharan-Africa migrants

Deborah N. Oluwasanya, PhD

## Abstract

Deception is best served as a sandwich - with just enough credibility to ensure the embedded falsehood is delivered without suspicion. AI-enabled deception via social engineering is one of the many risks that will remain an ongoing concern with the incorporation of AI into personal and corporate activities. Although the level of AI adoption remains unbalanced between the global north and south, migrants often serve as the underrated bridge to close the development gap. They are, however, often an underserved group with respect to digital safety and specifically tailored protective measures.

In this study, the vulnerability of Sub-Saharan-Africa (SSA) migrants to AI-enabled deception, using the risk of exposure to scam targeting, was evaluated. I hypothesized that the ability to distinguish human-generated content from AI-generated content had far-reaching implications beyond content assessment to determining vulnerability to AI-enabled deception. Data collected from a survey of 31 professionals and migrants from SSA across Europe and North America, covering themes on (i) Demographics and Transnational Context (ii) Core AI Literacy and Vulnerability, (iii) Mitigation and Trust, was modelled using a hybrid Structural Equation Model (SEM) and Multiple Linear Regression (MLR).

The results indicated that the strongest indicator of vulnerability to AI-enabled deception, such as scam, was prior exposure to targeting, as targeting was noted to be, in most cases, "a calculated attempt". Confidence in an individual's ability to identify AI content as well as the behavioral characteristics of high verification effort, emerged as significant protective factors that could lower the vulnerability to AI enabled deception. Other transnational contexts such as duration spent abroad or engaging in international fund remittance were found to have a small and insignificant effect on vulnerability. Based on results, policy recommendations are for intervention strategies to consider infrastructural protective measures, training that yields behavioral effects, and incidence reporting that allows for targeted support to lower the risk of scam targeting deteriorating into victimisation.

## Introduction

The democratisation of AI, such as the wide usage of Large language models and AI-generated content both on personal and corporate scale is a clear indication that the AI revolution, like the social media revolution, is here to stay. As much as there are clear benefits for using AI to lower the barrier of entry into aspects of creative communication, AI-enabled deception through social engineering is a clear and well-documented catastrophic risk which malicious actors can unfortunately apply in an infinite number of ways (Schmitt and Flechais 2024;

Bhardwaj 2025; Uplatz 2025). The awareness and usage of AI globally is heterogenous due to structural issues associated with digital access (World Economic Forum 2023, Microsoft AI Economy Institute 2026). So while some parts of the world are well aware of the potential of AI and the risks involved, others may not be so well informed.

Global mobility of people in many ways is associated with the distribution of value creation and migrants are often an underrated "bridge" through which the heterogeneous distribution of development is "smoothened" by the sharing of both knowledge and resources across different parts of the world with either similar or widely varying social-economic situations (Funds for NGOs, n.d; Cespedes-Reynaga 2025). This global mobility also necessitates the reliance on digital infrastructure to aid communication and resource transfer. Specifically on resource transfer via international fund remittances, Nigerian diasporan citizens have been shown to contribute as high as $24 billion in 2018, becoming the second most important source of foreign exchange (crude oil being the first), (Didia and Tahir, 2022), showing the significant utilization of financial digital infrastructure to enable this.

Prior to the widespread use of AI, cybersecurity has always been a concern and several jurisdictions have mechanisms to protect citizens from the harmful risk associated with malicious attempts (Manthovani 2023). Migrants relying on digital infrastructures are often an underserved group in this respect, as digital infrastructural solutions are not expressly tailored for them (Matlin et al., 2024), although they benefit from generic solutions that serve all users of digital infrastructure. It has been shown that Sub-Saharan Africa has the highest digital remittance costs globally (World Bank, 2025, Figure 13), signalling deep structural weaknesses in the region's digital financial infrastructure. This indicates that Sub-Saharan African migrants are systematically underserved by digital financial systems, a situation that may expose them to greater financial risks and reduced safety as they may seek alternatives to avoid high remittance costs.

The inclusion of AI-generated content into mainstream daily information consumption and AI's ability to generate realistic deep fakes that serve deception as a sandwich, with falsehood embedded in between just enough truth to be convincing and receivable, further accentuates cybersecurity vulnerabilities. Evidence of 'high-fidelity' deception is mounting, with recent high-profile incidents involving deepfake videos used to impersonate influential Nigerian leaders across the religious and commercial sectors (Leadership News, 2025; Awosika, 2026). Such incidents demonstrate the evolving capability of AI to bypass traditional visual verification methods used by immigrant communities.

In this study, the vulnerabilities of migrants from Sub-Saharan Africa to the AI catastrophic risk of deception were explored from the perspective of their ability to distinguish AI-generated content from human-generated content, while considering potential risk linkage to cross-border communication and financial transactions. This risk linkage is critical, as smartphones are becoming the primary channel for cross-border digital remittances to Sub-Saharan Africa. With mobile money adoption now reaching 40% of adults (World Bank, 2025, as cited in CNBC Africa, 2025), migrants' reliance on digital financial systems may increase their exposure to financial risks where infrastructure is weak. This study explored how behavioural and contextual characteristics could influence vulnerability and what factors served as protective indicators towards this risk.

## Materials and Methods

A ten-question survey (Appendix 1A) was drafted with the assistance of Copilot that covered 3 key areas: (i) Demographics and Transnational Context (ii) Core AI Literacy and Vulnerability (iii) Mitigation and Trust. The survey was then distributed digitally via direct messaging on LinkedIn and WhatsApp to friends and colleagues with a request to help pass the survey on to others within the demography of interest. All survey participants were known to be working professionals educated at least to a Bachelor's degree, with several having advanced degrees. Survey responses were collected over a 4-week period (Jan 6 - Feb 1, 2026). Survey recipients were contacted only once with no follow-up to avoid identifying recipients who responded to the survey directly, although some recipients confirmed completion of the survey. A total of 31 survey responses were collected and analysed as part of this study (Appendix 1B).

## Data Analysis

Before analysis, Survey response data was processed and cleaned, particularly for typographical errors associated with open-text responses. Graphical descriptions of survey respondents were generated using Microsoft Excel. A hybrid Structural Equation Model (SEM) and Multiple Linear Regression model was constructed using a cloud version of R on https://posit.cloud/ (Script Provided in Appendix 1C). Key packages included *tidyverse* for data wrangling, *lavaan* for SEM, and *ggplot2* for visualisation. SEM was chosen to enable the modelling of combined survey responses that generated latent, composite and derived variables (See table 1) while the Multiple Linear Regression model was constructed as a complement to the SEM to observe the behaviour of directly measured variables from the survey, relative to one another and for formulating the risk indicator model with policy implications. Model output tables were provided in Appendices 3 and 4. Graphs associated with relational analysis were generated.

## Use of AI in survey design and R-code scripting

Gemini AI was used in the initial ideation to develop survey questions and facilitated a search to ensure no overlap with other studies. The final survey questions were reviewed and decided upon by the author. In the same vein, Copilot was used to assist in writing code as well as in debugging errors generated by R- Script. Decisions on the direction of analysis and relationships to explore were those of the author.

# Results

## Demography of Survey Respondents

Survey respondents were primarily from Nigeria (97%) and comprised working professionals primarily resident in the United Kingdom (57%) and the United States (33%). Survey respondents were reasonably settled in their country of abode (majority had lived at least one year abroad at the time of survey), and a large proportion reported frequently remitting funds to their country of origin, possibly to support friends and family. Nearly a quarter had been targeted by AI-enabled deceptive scam attempts in the last 12 months. (Figure 1 A-D). The overwhelming majority reported using a smartphone as the primary device for accessing internet-based communication. This factor was therefore excluded from the analysis as there was no variation in the responses. Survey Questions and Responses collected are as reported in Appendix 1

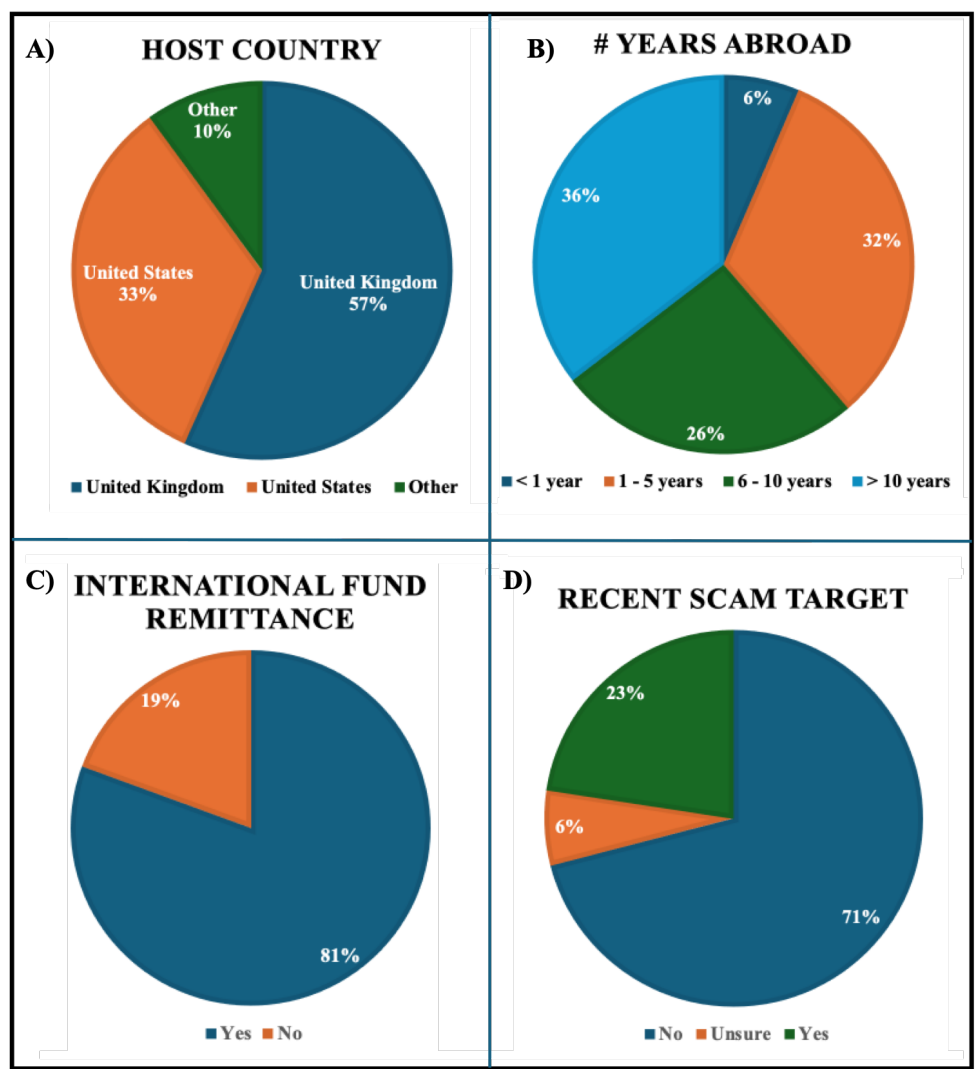

*Figure 1:* Demography of Survey respondents showing (A) Current County of residence (B) Number of Years away from country of origin (C) Proportion that remit funds frequently to home country (D) Proportion targeted by AI-enabled scam in the last 12 months.

# Definition of Latent, Composite and Derived Variables

Based on respondents' answers to the questions in the survey, latent variables (properties that were difficult to measure but could be inferred from a combination of observed variables) and composite variables (obtained by the combination of observed variables) were calculated to model factors that were likely to impact migrants' vulnerability to the AI risk of deception. Latent and Composite Variables used in this study, and their method of derivation, are summarised in the table below:

*Table 1: Definition of Latent, Composite and Derived Variables*

| Variable kind | Variable | Definition | Estimation |
|---|---|---|---|
| Latent | AI Literacy | Perceived ability to understand/detect and appropriately trust AI outputs | Estimated as a function of AI confidence (the ability to distinguish AI content from human content) and AI trust. AI Confidence was the dominant variable |
| | Transnational Exposure | Degree of ongoing cross-border engagement and financial ties | Estimated as a function of # years abroad and fund remittance activity |
| Composite | Institutional Trust | Trust in formal institutions (police/government) | Estimated as a function of trust in policy and trust in government institutions as an information source. *was modified in model to be limited to trust in policy due to model fit |
| | Vulnerability Index | Overall susceptibility combining experiential, cognitive, and behavioral risk | Estimated as a function of worry about scams; recent target of scam; the inverse of AI confidence and inverse of level of effort invested to verifying mischievous messages. |
| Derived | Observed verification behavior (verify_ord) | Behavioral protection: effort of verification of validity of mischievous messages | Recoded responses into ordinal scale (e.g., 0 none;1 some;2 high). Higher score = more verification (protective) |
| | AI confidence inverse | Cognitive vulnerability measure | Inverse of original scale (so higher = less confident); |

## Vulnerability Index as a Function of AI Confidence and Prior Exposure to Scam Attempts.

Results showed a negative relationship between Vulnerability index score and AI confidence (the ability to distinguish between AI-generated content and non - AI-generated content), suggesting that this ability to identify AI-generated content had a protective effect in the event of an AI-enabled deceptive attack (Figure 2A). On the other hand, the vulnerability score increased with exposure to scam targeting, which could imply that scam attempts are typically a calculated effort based on perceived vulnerability by malicious actors. The uncertainty around exposure to scam attempts in itself posed a risk to vulnerability, as reflected in higher scores than those who had not been a prior target. The low vulnerability score of individuals without prior scam exposure may be considered as a baseline vulnerability index that may change with exposure of individuals, except there are other known protective measures in place that make this group safely off the radar of potential scam targetters (Figure 2B). AI confidence showed an additive effect with prior scam exposure for individuals who reported that they were certain to have been targeted by scam or not. Both groups had different vulnerability index baseline scores (~ 3 for those with prior exposure and ~1 for those with none).

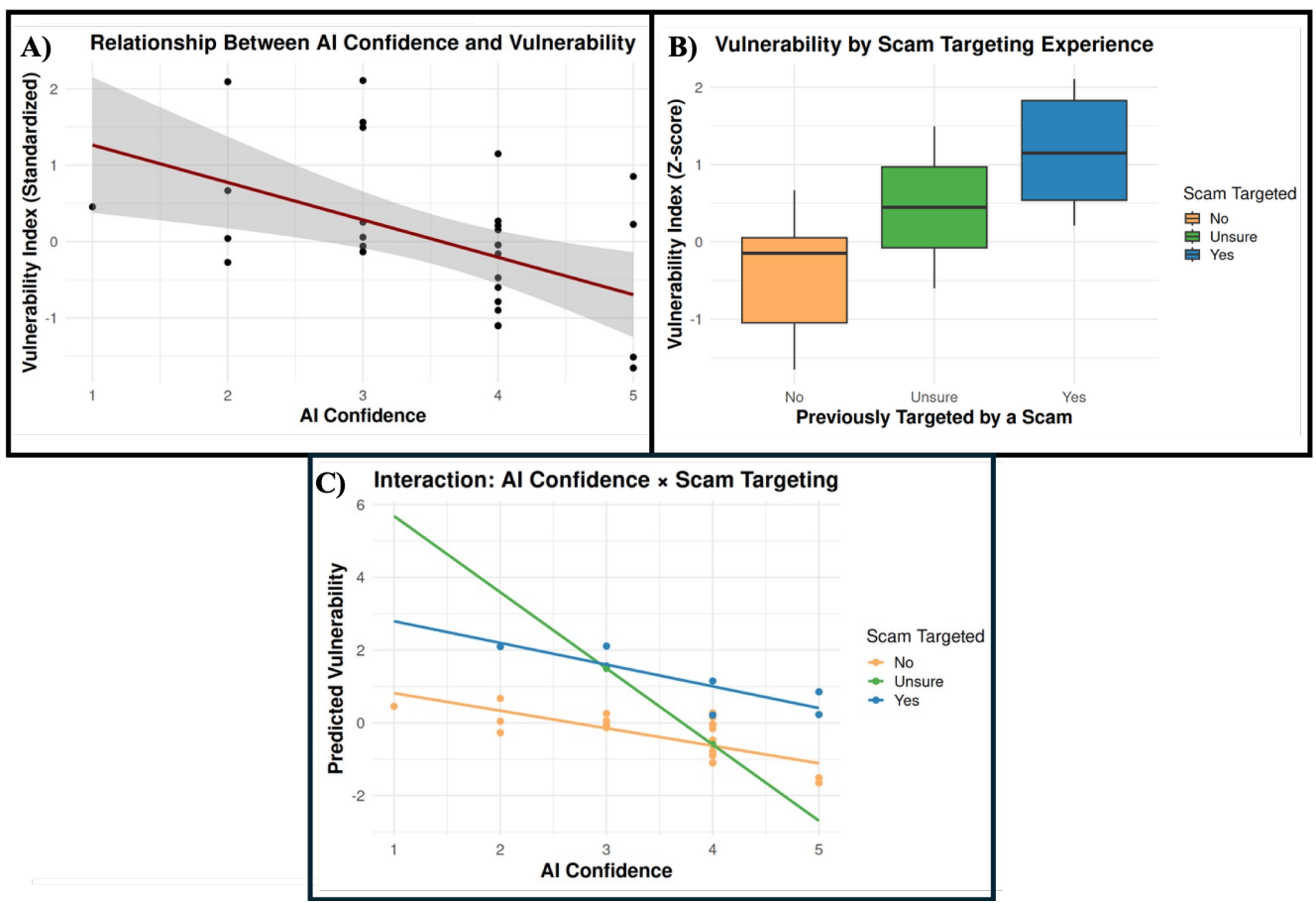

*Figure 2: Relationships Among AI Confidence, Scam Targeting Exposure, and Vulnerability Index Score (A) Negative association between AI Confidence and Vulnerability. (B) Distribution of Vulnerability across Scam Targeting experience groups, (C) Interaction between AI Confidence and Scam Targeting predicting Vulnerability.*

However, AI confidence imposed a protective measure on both categories to similar magnitudes relative to the initial baseline vulnerability index score. Respondents who were unsure that they had been the targets of recent scam attempts show the strongest reduction in their vulnerability index scores with an increase in AI confidence, suggesting that with increased AI confidence, they may be better able to identify future scam targeting and thereby reduce their vulnerability index score. Overall, AI confidence and prior scam exposure attempts were shown to have a significant interaction in influencing an individual's vulnerability index score (Figure 2C). The relationship between vulnerability index scores and other factors observed from survey responses was shown to be weak (Appendix 2). Worthy of note is the fact that time spent abroad has a very weak relationship with both vulnerability index scores and AI confidence (Appendix 2A and 2B), suggesting that time spent away from home country (as a proxy for better access to, and familiarity with AI-generated content, possibly arising from integration of AI in the work place) was not necessarily an indicator of how well a person could identify AI-generated content or a protective measure on a person's vulnerability index score.

## Evaluating Vulnerability to AI enabled Deception Using Structural Equation Modelling

A hybrid structural equation model was constructed using 30 of the 31 survey responses (1 response was excluded due to missing data on key variables). The model combined latent constructs and observed composites to examine predictors of vulnerability to AI-enabled deception. (Table 1). The overall model is summarised in the equation below.

Vulnerability_index_z ~ AI_Literacy + Transnational_Exposure + Institutional_Trust + scam_num

The model was assessed to fit suitably, explaining 83% of the variance observed (details of model fit and output table are as in Appendix 3). Of all variables, Scam_num - representing prior exposure to scam targeting emerged as the strongest and statistically significant predictor ($\beta = 0.86$, $p < .001$) relative to all other factors considered. AI Literacy on the other hand, although not statistically significant (possibly due to small sample size), also showed a large effect size ($\beta = -0.62$, $p = .141$). While every 1 unit of scam exposure predicted a 0.86 unit increase in vulnerability score, every unit increase of AI Literacy (particularly AI confidence, per model estimators), predicted a 0.62 reduction in vulnerability index score.

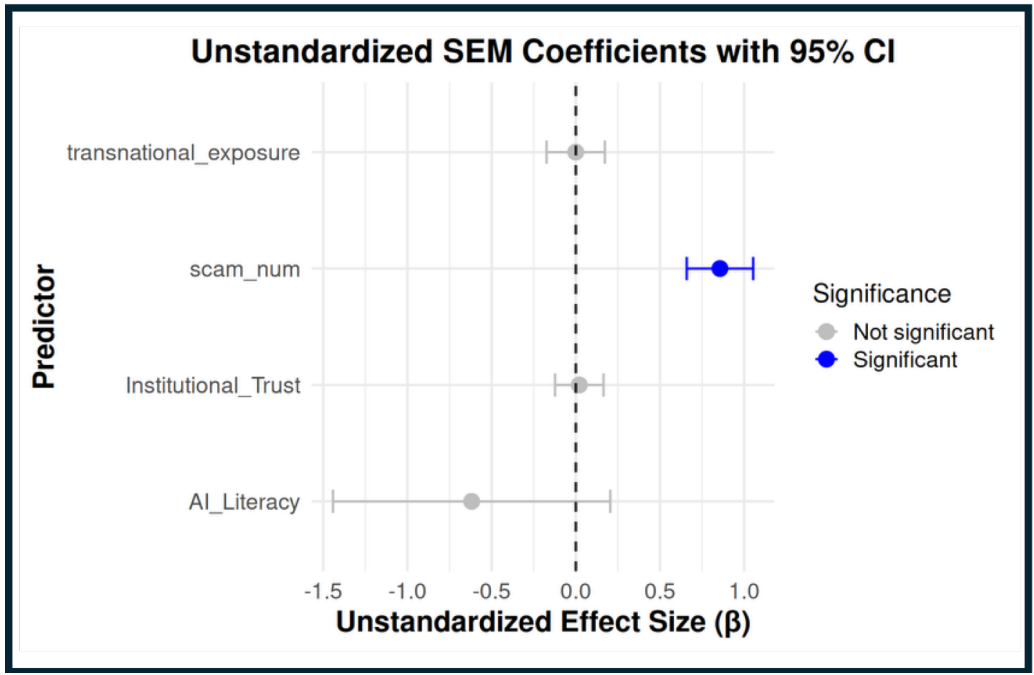

*Figure 3:* Overview of Effect sizes and Statistical significance of SEM model predictors on Vulnerability.

## Evaluating Risk Indicators Using Multiple Linear Regression

To complement the hybrid SEM, a risk indicator model was estimated using only observed, directly interpretable variables to identify which measurable characteristics are most strongly associated with vulnerability. A linear regression was fitted based on the equation below
vulnerability_index_z ~ ai_conf_inv + trust_police + years_abroad_ord + remit_cat + scam_targeted_cat + verify_ord

The risk indicator model was suitably fitted such that 88% of the variance was explained by the model (full model output is provided in Appendix 4). Verification behaviour coded as verify_ord ($\beta$ = - 0.67, $p < .001$), Prior Scam targeting coded as Scam_targetted_CatYes ($\beta$ = 1.31, $p < .001$), AI confidence inverse coded as ai_conf_inv ($\beta$ = 0.37, $p < .01$) emerged as statistically significant with verification behaviour and AI confidence (Inver form used in model implies, high values have a lowering predictive effect ) indicating protective effect, lowering vulnerability score. Years abroad and remittance activities did not have any significant effect on predicted vulnerability, indicating that these factors were independent of susceptibility to AI-enabled deception, suggesting that migrants are not particularly predisposed to this risk due to their need to remit funds abroad.

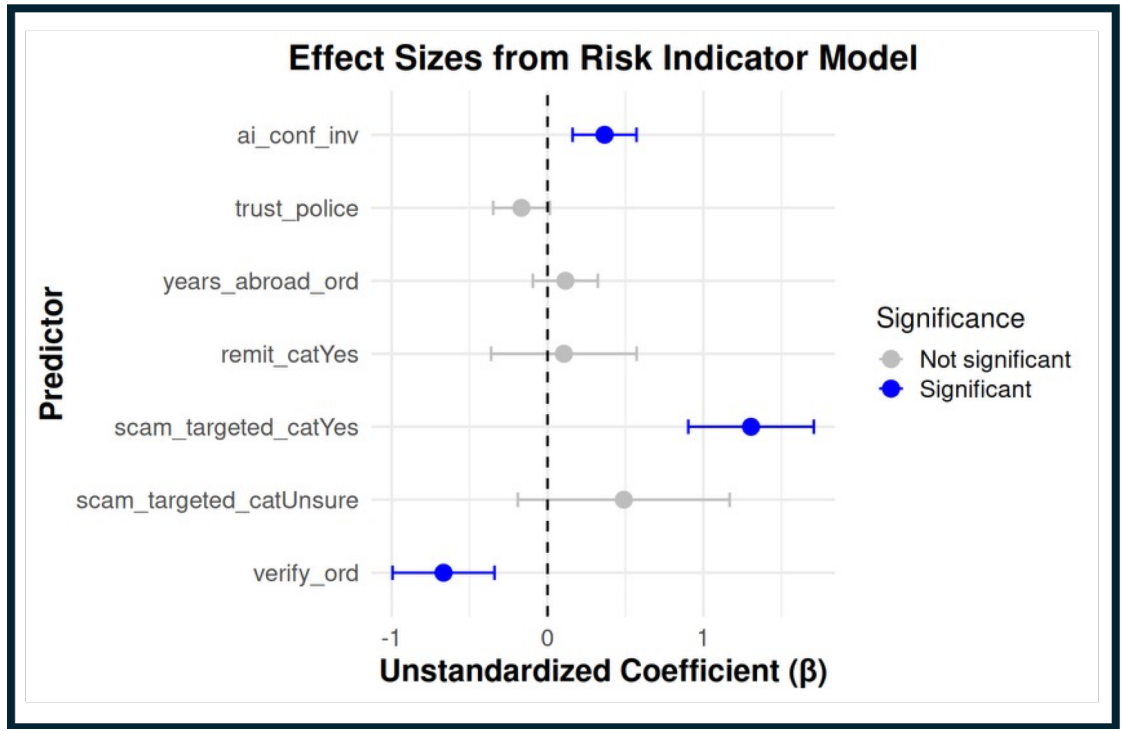

*Figure 4: Overview of Effect sizes and Statistical significance of Risk Indicator model predictors on Vulnerability.*

## Implications of Scam Exposure, AI confidence and Verification Behaviour on Vulnerability

Scam Exposure is very unlikely to be an isolated one-off attempt, and risk is elevated with every exposure. This data indicates that efforts aimed at preventing exposure to scam events will be valuable. As vulnerability is a combination of several interacting factors, whereas this study may not have fully captured the breadth of activity of individuals, making them more predisposed to this risk - see Figure 2 for further details.

The protective effect of verification behaviour (coded from responses ranging from zero to low and high effort) in reducing vulnerability implies the potential of verification behaviour to serve as a modifiable skill with real impact, allowing individual verify information sources before acting on it, thereby lowering the susceptibility to AI deception risk (Figure 5b).

As noted previously, AI confidence also had a protective effect (Figure 5c shows inverted AI confidence, where lower scores correspond to higher AI confidence and lower predicted vulnerability), with implications for determining the capability of individuals to assess the credibility of information that will inform their response accordingly.

Taken together, interventions to prevent scam exposure and training to enhance confidence to assess the credibility of information that then leads to behavioural attributes that invest high effort to verify information before taking action will contribute significantly to lowering the vulnerability of individuals to AI deception risks.

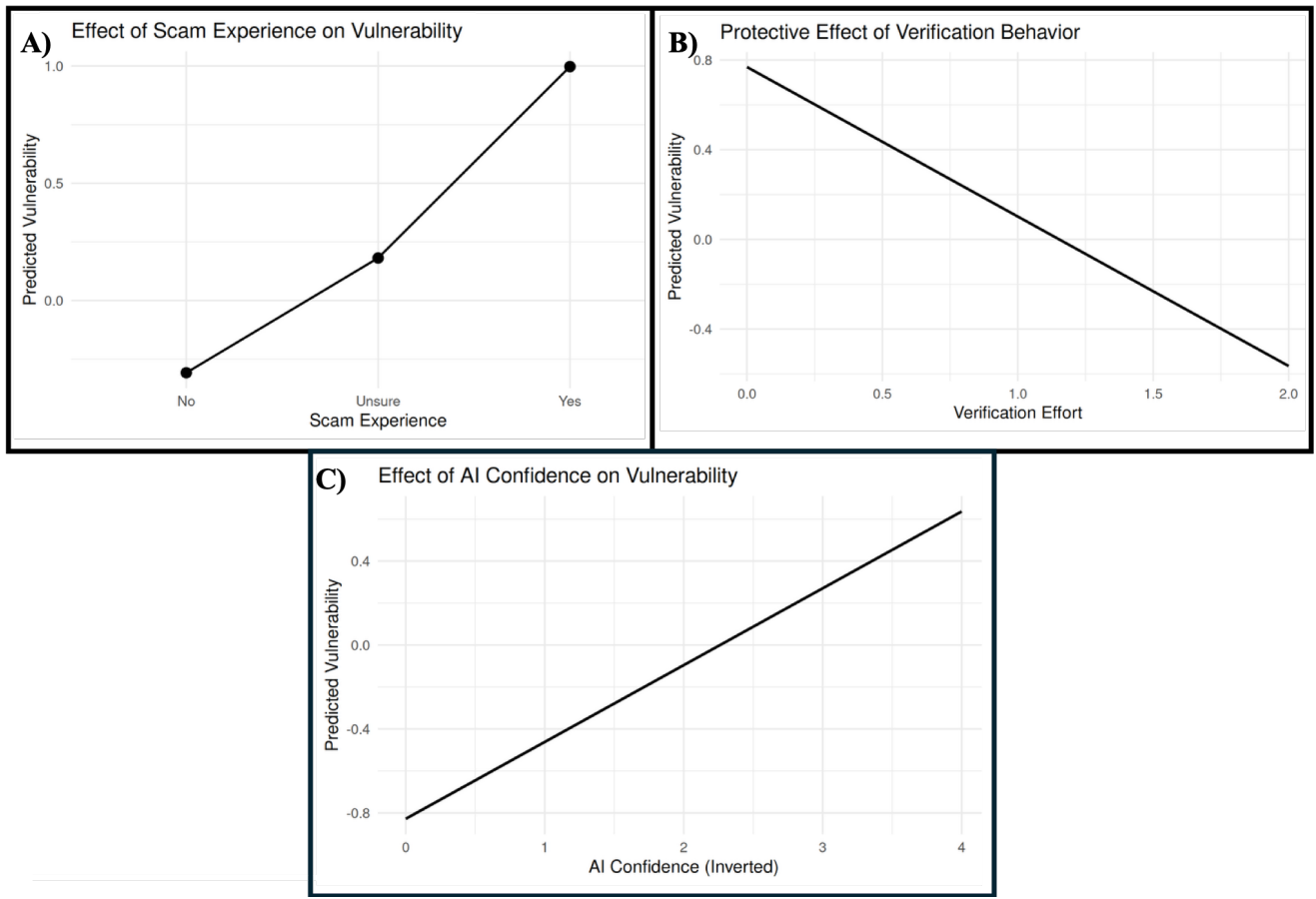

*Figure 5*: Key Predictors of Scam Vulnerability in the Risk-Indicator Model (A) Prior Scam Experience (B) Verification Behaviour (ranging from no effort to high effort) (C) Inverse of AI confidence - where lower scores indicate higher confidence and thus lower vulnerability.

## Discussion

This study did not identify a relationship between the fund remittance activity of SSA migrants internationally and high vulnerability to AI-enabled scam, possibly due to financial transactions being a regular part of life and digital infrastructure enabling international funds transfer seemingly not having a risk profile different from local daily financial transactions and requirements to remit funds locally. Cyber scam is already a well known risk that has possibly enabled financial digital infrastructure to invest adequately towards ensuring safety on an ongoing basis (Adejumo and Ogburie 2025). However the major risk lies not in the vulnerability of infrastructure but in the vulnerability of individuals to social engineering enabled by AI deception. A previous study identified "Overconfidence" and "Social Influence" (defined in study as ease with which a person is persuaded to act) as significant positive predictors of vulnerability to online fraud (Balkrishna et al., 2025). A deeper understanding of the factors that make individuals more likely to be targeted by scammers is essential for uncovering how AI-enabled deception may be strategically deployed to heighten victim vulnerability. This study found a strong association between vulnerability and prior exposure to AI-enabled scams, suggesting that scam targeting is likely a calculated effort rather than a random occurrence. As noted by Langenderfer and Shimp (2001), deceptive actors initially provide a bait to which targets self-select by responding, enabling deceptive actors evaluate their potential as candidates for targeting. The mechanism for evaluation deceptive actors deploy is yet unclear but a systematic review by Norris et al.,(2019) identified three factors common to all baiting processes - the message, experience and disposition. What is more, Titus and Grover (2021) already established that repeat victimisation with respect to fraud and by implication, repeat targeting is highly prevalent.

In this study, AI literacy (as measured by the self-reported ability of respondents to identify AI-generated content and awareness that Large Language Models are not an unbiased source of truth) and behaviour marked by the degree of effort invested into verifying information received are two potent factors that could provide protective effect on individuals relative to vulnerabilities to AI-enabled deception. Verification behaviour as a protective factor with respect to fraud victimisation was previously established by Titus and Grover (2021). AI Literacy would empower individuals to assess the credibility of information provided to them, while verification behaviour will equip individuals with the recognition of the need to invest in testing, examining and cross-checking/fact-checking information before taking action relative to information exposed to in a broader sense. Investment of efforts into AI literacy would therefore be most impactful on safety from AI-enabled deception when knowledge is translated into behaviours. A further study into the interaction between AI literacy and Behavioural change and how each impacts one another, if there is a feedback loop will be a valuable direction to pursue. Prior to this an actionable step will be training efforts that leads to behavioural change such as an "innoculation" type training experimented with by Robb et al., (2023) in other contexts for which in our context, individuals may be exposed to simulated AI deception attempts and training provided based on their responses to ensure adequate behaviour in the case of real AI-enabled deception risk exposure.

## Summary

This study investigated the vulnerability of Sub-Saharan African migrants who rely heavily on digital communication and cross-border financial interactions to AI-enabled deception, particularly scams that exploit transnational ties. Using a hybrid Structural Equation Model (SEM) and a Multiple Linear Regression risk-indicator model, the study identifies the behavioural, cognitive, and contextual factors that shape vulnerability.

## Key Findings

- Prior scam exposure is the strongest predictor of vulnerability.
  Scam targeting is not random; individuals who have been targeted once are significantly more likely to be targeted again.

- AI confidence (ability to distinguish AI-generated from human content) is protective. Higher AI confidence predicts lower vulnerability, even after accounting for scam exposure.
- Verification behaviour is an effective, modifiable protective factor.
Individuals who invest effort in verifying suspicious messages show substantially lower vulnerability scores.
- Transnational factors (years abroad, remittance activity) do *not* predict vulnerability. Migrants are not inherently more vulnerable because they remit funds or live abroad; the risk lies in social engineering, not financial infrastructure.
- Institutional trust and AI trust have weak or non-significant effects.
Vulnerability is driven more by behavioural and cognitive factors than by trust in authorities and primary source of information.

## Policy Recommendations

Policies relating to infrastructure
- Governments should require the declaration of, or labelling of AI-generated content released into the public domain.
- Communication platforms (messaging apps, social media, etc) should incorporate deepfake/AI-generated content detection to protect users from misinformation and deception
- Financial institutions should expand fraud protection investment to cover AI-enabled social engineering.
- Regulatory bodies should require the integration verification prompts into messaging apps and remittance platforms

Policies relating to behaviour
- Investment in AI Literacy efforts should be those that results in behavioural change
- If we consider AI-enabled deception as a virus, then "innoculation" style training that exposes individuals to realistic deception attack will enable tailored training on behavioural tendencies to enhance protective behaviours
- Policies should encourage "pause-and-verify" habits before engaging with, or acting on digital information from any source.

Policies relating to Exposure
- Embassies and host countries should partner in developing a joint incidence reporting platform - in similar fashion employed in the aviation sector
- Protective intervention strategies (behavioural and infrastructural) should be instituted upon first reporting so as to avoid risk of targeting yielding actual victimization.
- Informal community groups should be equipped with training to enable the provision of first line defence support to targeted individuals.

# Conclusion

Vulnerability to AI-enabled deception among SSA migrants is shaped primarily by exposure history, AI literacy, and verification behaviour rather than by demographic or transnational characteristics. This suggests that targeted behavioural interventions could meaningfully reduce risk.

# Appendix 1

A) AI Literacy and Transnational Vulnerability amongst SSA migrants in the West - Survey Questions: https://forms.gle/ufx3G6bPqqvzWShDA

Section 1: Demographics and Transnational Context (3 Questions)

1. Country of Origin/residence:
    - What country in Sub-Saharan Africa do you consider your country of origin? (Open-text)
    - What country do you currently reside in? (Open-text)
    - How long have you been away from your home country? (multiple choice answer with options - less than 1 year, 1 - 5 years, 6 - 10 years, and over 10 years)
2. Remittance Activity:
    - Do you regularly send money (remittances) to family members or contacts in your country of origin? (Yes / No)
3. Digital Access/Usage:
    - Which device do you use most often to access news, information, and communication? (Select one: Smartphone, Laptop/PC, Other)

---

Section 2: Core AI Literacy and Vulnerability (4 Questions)

4. AI Literacy and Accuracy (Combined Question including Hallucination):
    - Please rate your agreement with the following two statements: (Likert Scale: 1-Strongly Disagree to 5-Strongly Agree)
        - A. I am confident in my ability to distinguish content (audio, video, or text) created by AI from content created by a human.
        - B. If an AI chatbot (like ChatGPT) provides an answer with confidence, I can trust it is factually correct. (Measures awareness of AI Hallucination/Inaccuracy)
5. Exposure to Transnational Scamming:
    - In the past 12 months, have you been targeted by a scam attempt that specifically mentioned or involved a family member, friend, or contact in your country of origin (e.g., claiming a relative is in distress)?
        - (Yes / No / Not sure)
6. Deception Modality (AI-Relevant Impersonation):
    - If you answered YES above, what was the method used to impersonate the person you know? (Select all that apply)
        - Text message/Email
        - Voice message/Call (using their *actual voice* but edited/cloned)
        - Video call/Photo (using their *actual image* but edited/deepfake)
        - The scam did not involve impersonation.

7. Vulnerability & Concern Statement:
   - Please rate your agreement with the following statement: *"I worry about my transnational identity and family ties making me a target for sophisticated online scams."*
     - (Likert Scale: 1-Strongly Disagree to 5-Strongly Agree)

---

Section 3: Mitigation and Trust (3 Questions)

8. Verification Action (Open-Text):
   - When you receive a suspicious message that appears to be from a relative or friend, what is the *very first thing* you do to try and verify it? (Open-text response)
9. Trust in Authorities:
   - How much do you trust the police or other government agencies in your host country to help you if you were a victim of a transnational online scam?
     - (Likert Scale: 1-Do not trust at all to 5-Trust completely)
10. Preferred Information Source:
    - Which of the following sources would you trust most to provide accurate information on new AI-related scams? (Select one)
      - Official host country police/government websites.
      - Community leaders/religious organizations.
      - Social media/WhatsApp groups from my diaspora community.

B) Survey Raw Responses:
https://docs.google.com/spreadsheets/d/1d4-Pm6AzfgiyL9XssbQ18AiKOV6z_jR0UlDzwuGglSY/edit?usp=sharing

C) Data Analysis Script in R
https://drive.google.com/file/d/1Io3wkvNpEi0R2yy9_BWcBM19q5BXYNDj/view?usp=sharing

# Appendix 2

Relationship Between Vulnerability Index, Institutional Trust, Numbers of Years resident abroad (reported as an ordinal factor) - All relationships were noted to be weak suggesting independence of factors from one another.

A)

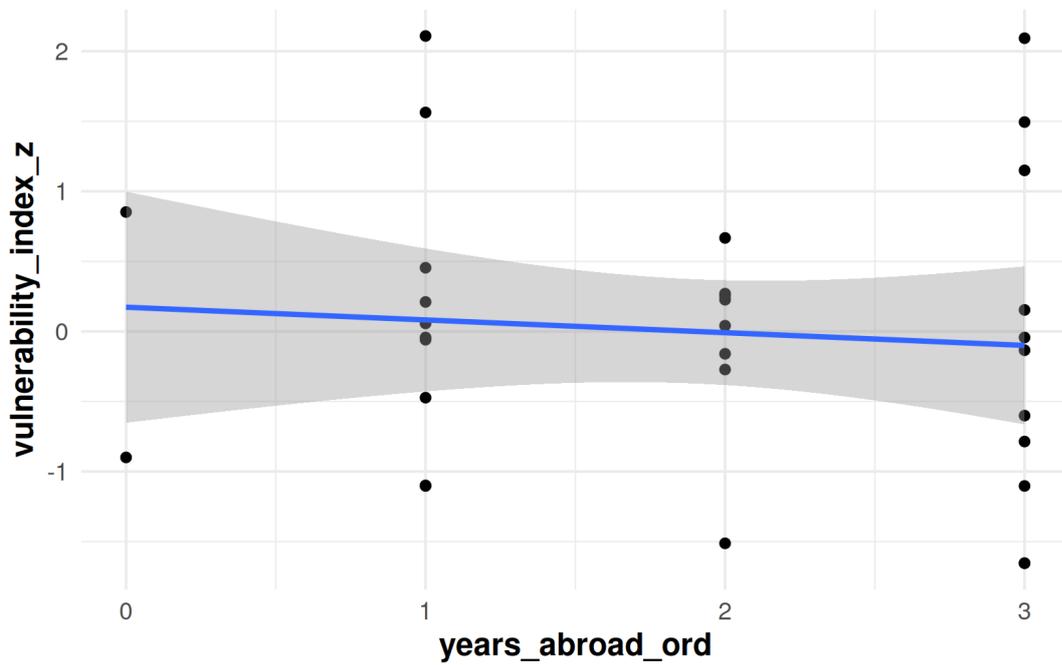

B)

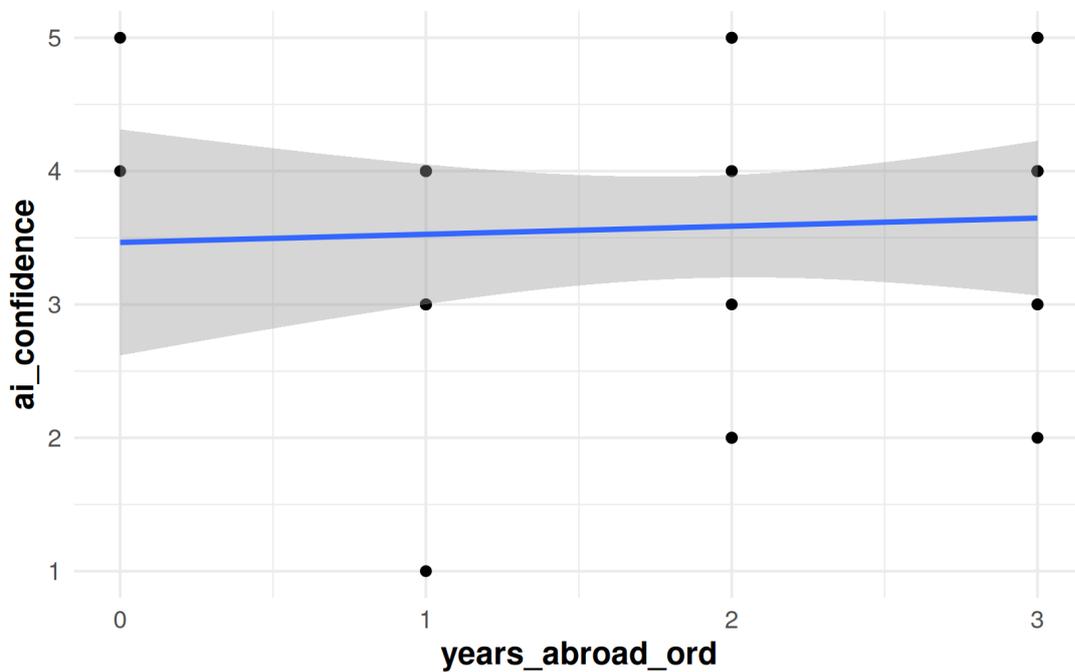

C)

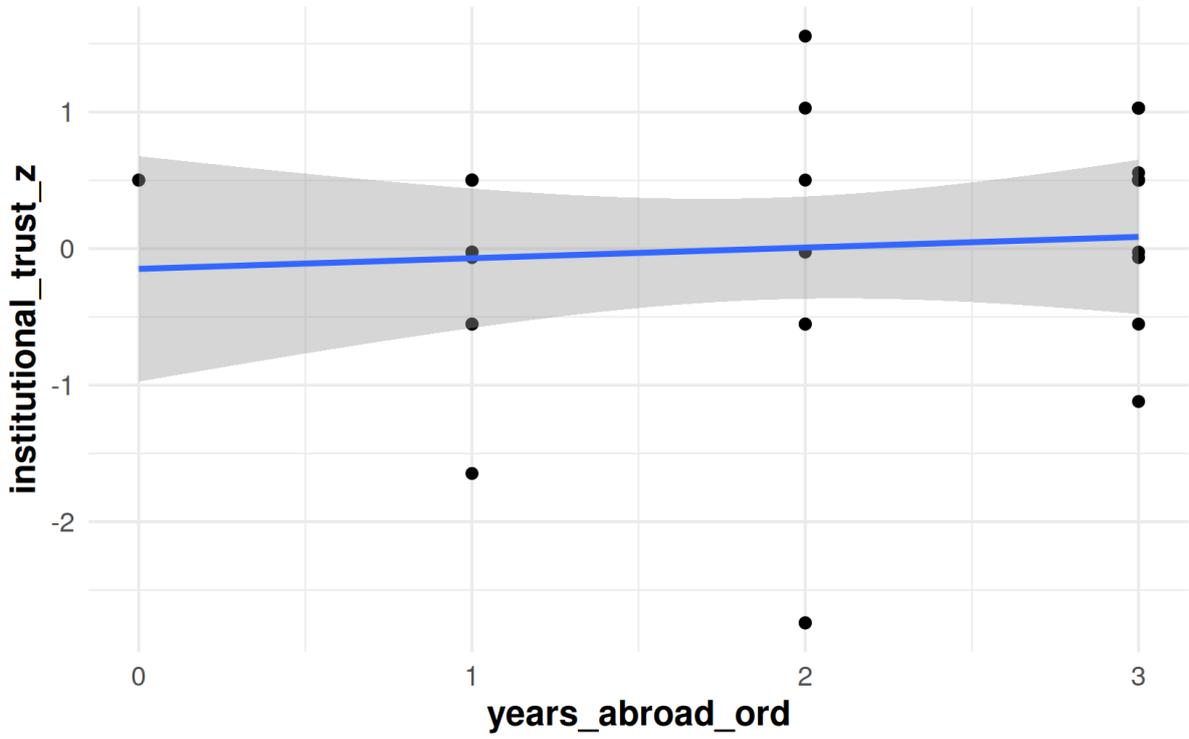

D)

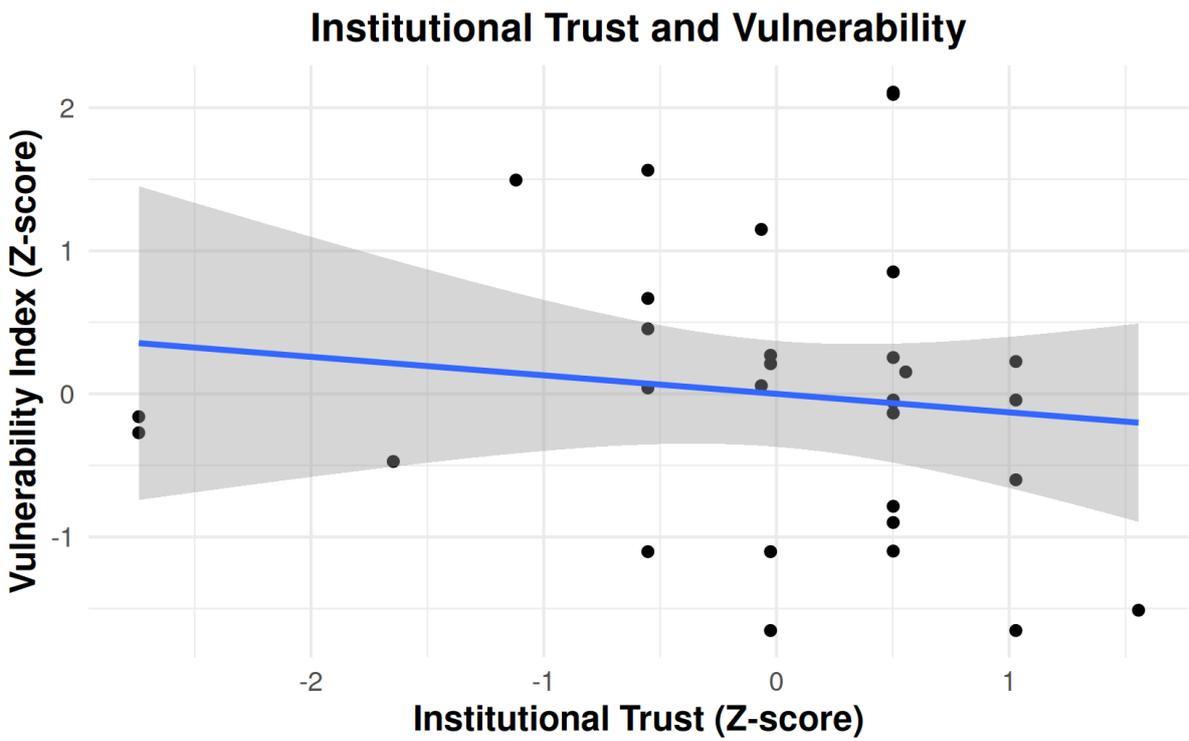

E)

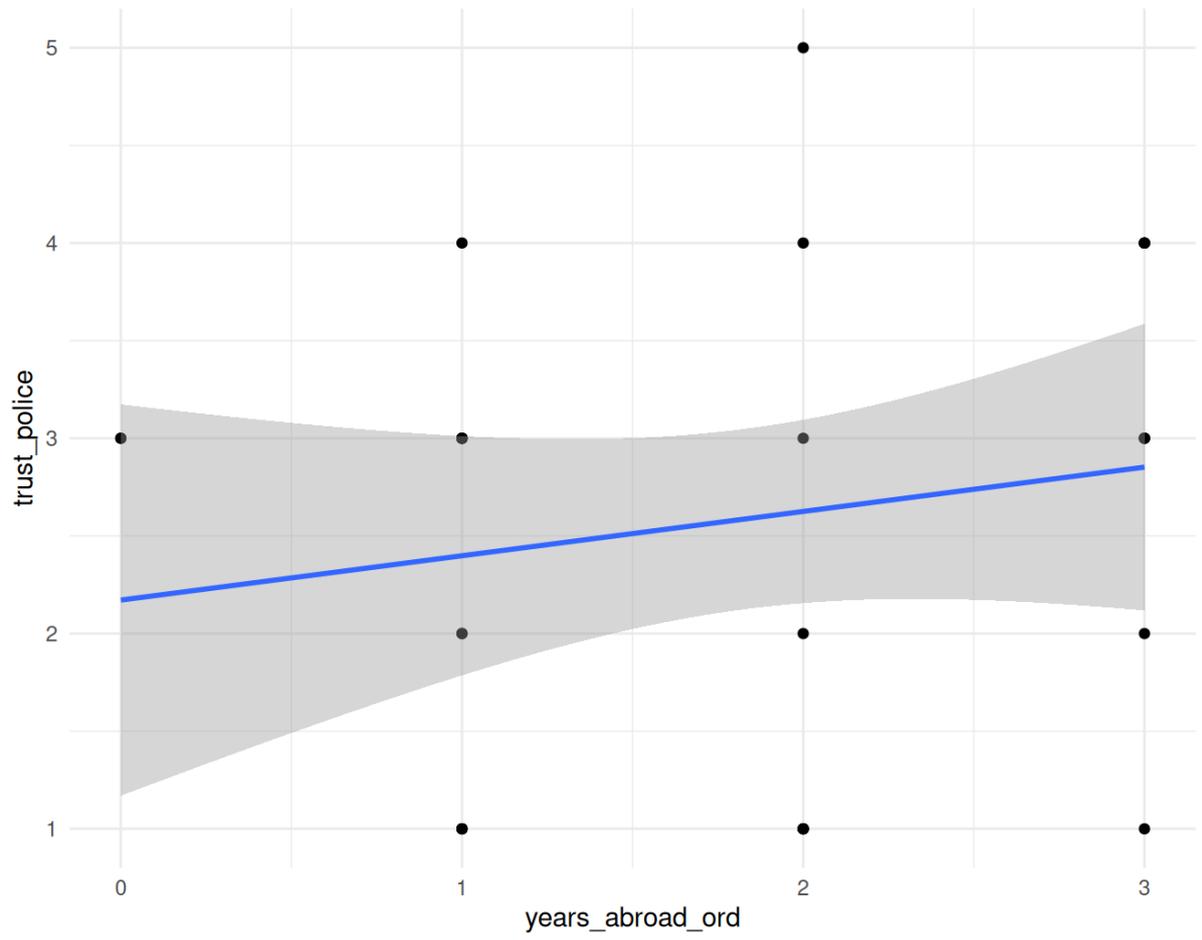

# Appendix 3

Structural Equation Model Results

Model Fit Indices

| Fit Index | Value |
|---|---|
| Chi-square (df) | 9.86 (7) |
| p-value | .197 |
| CFI | .934 |
| TLI | .869 |
| RMSEA | .117 |
| RMSEA 90% CI | [.000, .270] |
| p(RMSEA ≤ .05) | .240 |
| p(RMSEA ≥ .08) | .690 |
| SRMR | .125 |
| AIC | 328.65 |
| BIC | 344.06 |
| SABIC | 309.82 |
| Log-likelihood (H0) | −153.33 |
| Log-likelihood (H1) | −148.40 |

Measurement Model: Factor Loadings

| Latent Variable | Indicator | Estimate | SE | z | p | Std.lv | Std.all |
|---|---|---|---|---|---|---|---|
| AI Literacy | ai_confidence | 1.000 (fixed) | — | — | — | 0.972 | 0.950 |
|  | ai_trust | 0.227 | 0.237 | 0.959 | .337 | 0.221 | 0.211 |
| Institutional Trust | trust_police | 1.000 (fixed) | — | — | — | 1.174 | 1.000 |

Structural Paths

| Outcome | Predictor | Estimate | SE | z | p | Std.lv | Std.all |
|---|---|---|---|---|---|---|---|
| vulnerability_index_z | AI_Literacy | −0.619 | 0.420 | −1.473 | .141 | −0.601 | −0.582 |
|  | Instittnl_Trst | 0.020 | 0.073 | 0.279 | .781 | 0.024 | 0.023 |
|  | transntnl_xpsr | −0.001 | 0.088 | −0.008 | .994 | −0.001 | −0.001 |
|  | scam_num | 0.856 | 0.101 | 8.499 | <.001 | 0.856 | 0.700 |
| Institutional_Trust | transntnl_xpsr | 0.186 | 0.217 | 0.855 | .393 | 0.158 | 0.154 |

Variances

| Variable | Estimate | SE | z | p | Std.lv | Std.all |
|---|---|---|---|---|---|---|
| ai_confidence (resid.) | 0.102 | 0.628 | 0.162 | .872 | 0.102 | 0.097 |
| ai_trust (resid.) | 1.050 | 0.273 | 3.846 | <.001 | 1.050 | 0.956 |
| trust_police (resid.) | 0.000 | — | — | — | 0.000 | 0.000 |
| vulnerability_index_z (resid.) | 0.183 | 0.245 | 0.746 | .456 | 0.183 | 0.171 |
| AI_Literacy (latent var.) | 0.944 | 0.683 | 1.383 | .167 | 1.000 | 1.000 |
| Institutional_Trust (resid.) | 1.346 | 0.348 | 3.873 | <.001 | 0.976 | 0.976 |

# Appendix 4

Risk Indicator model (Multiple Linear Regression) Predicting Standardized Vulnerability Index (Six observations were removed due to missingness).

| Predictor | Estimate | Std. Error | t | p |
|---|---|---|---|---|
| Intercept | 0.45781 | 0.49360 | 0.927 | 0.3667 |
| AI Confidence (inverted) | 0.36611 | 0.09707 | 3.772 | 0.0015 ** |
| Trust in Police | −0.16677 | 0.08599 | −1.939 | 0.0692 · |
| Years Abroad (ordinal) | 0.11522 | 0.09874 | 1.167 | 0.2594 |
| Remits (Yes) | 0.10512 | 0.22128 | 0.475 | 0.6408 |
| Scam Targeted (Yes) | 1.30605 | 0.19089 | 6.842 | <0.001 *** |
| Scam Targeted (Unsure) | 0.48959 | 0.32191 | 1.521 | 0.1467 |
| Verification Effort | −0.66680 | 0.15505 | −4.301 | 0.0005 *** |

Model Fit

| Statistic | Value |
|---|---|
| Residual SD | 0.3879 |
| df (residual) | 17 |
| Multiple $R^2$ | 0.8773 |
| Adjusted $R^2$ | 0.8267 |
| F-statistic | 17.36 (df = 7, 17) |
| Model p-value | $1.36 \times 10^{-6}$ |

Significance codes: *** p < .001; ** p < .01; * p < .05; · p < .10.